\documentclass[12pt,titlepage]{article}

\usepackage{epsfig}
\usepackage{amsmath}
\usepackage{amssymb}
\usepackage{lscape}
\usepackage{hyperref}
\usepackage{cite}
\usepackage{amsmath,epsf,color,shadow,pifont}

\numberwithin{equation}{section} 
\numberwithin{figure}{section}
\numberwithin{table}{section} 
\begin{document}

\begin{titlepage}
\begin{center}

{\Large {\bf Universality in generalized models of inflation.}}
 \\
 ~\\
\vskip 2cm

\vskip 1cm

{\bf  \large P. Bin\'etruy${}^{a,b,}$\footnote{Member of the Institut Universitaire de France.}\footnote{pierre.binetruy@apc.univ-paris7.fr}, J. Mabillard${}^{c,}$\footnote{joel.mabillard@ed.ac.uk}, M. Pieroni${}^{a,b,}$\footnote{mauro.pieroni@apc.in2p3.fr}. }\\
~\\
~\\
{${}^{a}$\em AstroParticule et Cosmologie, Universit\'e Paris Diderot, CNRS, CEA, Observatoire de Paris, Sorbonne Paris Cit\'e,
10, rue Alice Domon et L\'eonie Duquet, F-75205 Paris Cedex 13, France} \\
{${}^{b}$\em Paris Centre for Cosmological Physics, Universit\'e Paris Diderot,
10, rue Alice Domon et L\'eonie Duquet, F-75205 Paris Cedex 13, France}\\
{${}^{c}$\em School of Physics and Astronomy, University of Edinburgh, Edinburgh, EH9 3JZ, United Kingdom}\\

\end{center}

\vskip 1cm
\centerline{ {\bf Abstract}}{ We discuss the cosmological evolution of a scalar field with non standard kinetic term in terms of a Renormalization Group Equation (RGE). In this framework inflation corresponds to the slow evolution in a neighborhood of a fixed point and universality classes for inflationary models naturally arise. Using some examples we show the application of the formalism. The predicted values for the speed of sound $c_s^2$ and for the amount of non-Gaussianities produced in these models are discussed. In particular, we show that it is possible to introduce models with $c_s^2 \neq 1$ that can be in agreement with present cosmological observations.}

\vskip .5cm
\indent

\vfill

\end{titlepage}

\newpage

\tableofcontents

\section{Introduction.\label{sec:introduction}}
The $\Lambda$CDM model is widely accepted as the theory to describe the evolution of our Universe. Its success is due to the high concordance of its predictions with experimental observations in spite of its small amount of parameters. While this model presents some theoretical problems related with its early time behavior their solutions seem to be naturally furnished by inflation. Even if nowadays the main mechanisms of inflation begin to be almost clear, the definition of a concrete model still bears several issues. A wide set of models has been proposed over the years and in some cases theoretical predictions are so close that they are nearly indistinguishable. This suggests that it can be both interesting and instructive to analyze the problem from a different point of view and also emphasizes the importance of finding some universality among all the models. \\

In order to have a better connection between theory and experimental observations, we have proposed in~\cite{Binetruy:2014zya} a classification of inflationary models that relies on the simplest property of inflation i.e. the nearly exact scale invariance. Inspired by a formal similarity, it is possible to describe inflation using a renormalization group equation (RGE) expressed in terms of a beta function similar to the one that is well known in quantum field theory. In this framework it is thus possible to describe inflation using the Wilsonian picture of renormalization group (RG) flows between fixed points. From this point of view, it is easy to identify the terms in the equations of motion which drive inflation and in particular it is straightforward to classify inflationary models in terms of a minimal set of parameters. A major benefit of this formalism is the possibility of defining universality classes, that similarly to the ones defined in statistical mechanics, should be intended as sets of theories sharing a single scale invariant limit. \\

Strong motivations to support the introduction of this formalism are offered by the idea of applying holography to cosmology~\cite{McFadden:2009fg,McFadden:2010na}. This picture has started to spread over the last years~\cite{Garriga:2014fda,Garriga:2014ema,Garriga:2015tea} and indeed allows to find intriguing interpretations for different cosmological topics and in particular for inflation~\cite{Kiritsis:2013gia}. Interestingly, in the holographic description the departure from a de Sitter (dS) configuration (\emph{i.e.} inflation) is interpreted as a departure from conformal invariance of the dual (pseudo) Quantum Field Theory (QFT). The departure from conformal invariance is described by an operator $\mathcal{O}(x)$, dual to the inflaton field, that is characterized by a scaling dimension $\Delta$. The classification of these operators is equivalent to the classification of the different inflation theories. The holographic interpretation of some universality classes defined in terms of the $\beta$-function formalism for inflation has been discussed in~\cite{Binetruy:2014zya}, in~\cite{Kiritsis:2013gia} and in~\cite{Pieroni:2016gdg}. \\

While the analysis presented in~\cite{Binetruy:2014zya} suits perfectly the case of single field models with standard kinetic term and minimal coupling with gravity, the extension to generalized models of inflation is still uncompleted. In particular in this paper we focus our interest on models where the inflaton has a non-standard kinetic term. As it has been shown in~\cite{ArmendarizPicon:1999rj}, a non-standard kinetic term can drive an period of inflation, called K-inflation. Explicit and well-known examples are the Dirac-Born-Infeld (DBI) model~\cite{Silverstein:2003hf,Alishahiha:2004eh} or tachyon inflation~\cite{Gibbons:2002md,Padmanabhan:2002cp,Steer:2003yu} (for a comparison between the theoretical predictions of these models and observational constraints see for example~\cite{Li:2013cem}). A discussion of these models in terms of the Hamilton-Jacobi approach of Salopek and Bond~\cite{Salopek:1990jq} was carried out in~\cite{Garriga:2015tea}. In particular the authors have clearly pointed out that the evolution of the system can be consistently carried out in terms of a Renormalization Group Equation (RGE). \\

In the present paper we propose a slightly different method to discuss models with non-standard kinetic terms exploiting the Hamilton-Jacobi formalism. In particular, we propose a different definition for the $\beta$-function that enforces the invariance of $\beta$ under reparametrizations of the inflaton field. This different definition is inspired by the analysis carried out in~\cite{Pieroni:2015cma}, where the formalism was generalized in order to discuss models where the inflaton is non-minimally coupled with gravity. In particular, it has been shown that the Einstein frame formulation of models where the inflaton is non-minimally coupled with gravity also admits a natural description in terms of the $\beta$-function approach. Since the Einstein frame formulation of these models is equivalent to a subclass of models with non-standard kinetic terms, the analysis presented in this paper is a generalization of the formalism which goes beyond (and indeed incorporates) the analysis presented in~\cite{Pieroni:2015cma}.\\

The outline of this paper is the following: In Sec.~\ref{sec:non-standard} we start by giving a brief review of the main characteristic of inflationary models with non-standard kinetic term. In particular we point out some fundamental differences with the standard case. In Sec.~\ref{sec:formalism} we present the generalization of the $\beta$-function formalism to models with non-standard kinetic terms. Some useful formula are reported in Appendix~\ref{sec:appendix_formula} and some more theoretical arguments are presented in Appendix~\ref{sec:appendix_theory}. Sec.~\ref{sec:general} is dedicated to the study of some explicit examples and we show how the definition of a set of universality classes for inflationary models is carried out. Another interesting example (\emph{i.e.} a class of models which are inspired by tachyon inflation) is reported in Appendix~\ref{sec:appendix_tachyon}. Our concluding remarks are finally left to Sec.~\ref{sec:conclusions}.

\section{Non-standard kinetic terms: a brief review.\label{sec:non-standard}}
In models with generalized kinetic term, it is possible to go beyond the simplest realization of inflation \emph{i.e.} slow roll inflation. In particular in these models the phase of exponential expansion can be driven by the non-standard kinetic term of the inflaton field. In this section, we briefly review models with a non-standard kinetic term and we explain this mechanism.\\

We start by considering the most general expression for the action describing a homogeneous classical scalar field $\phi(t)$ minimally coupled with gravity in the Einstein frame\footnote{Working in a Friedman-Lema\^itre-Robertson-Walker with scale factor $a(t)$ and assuming for simplicity a flat metric in terms of cosmic time i.e. $\textrm{d}s^2 = -\textrm{d}t^2 +a(t)^2 (\textrm{d}r^2 + r^2 \textrm{d}\Omega^2)$.}:
\begin{align}
\label{eq_non_standard:action}
		S&=\int\mathrm{d}^4x\sqrt{-g}\left(\frac{R}{2\kappa^2} +p(\phi,X)\right) \ ,
	\end{align}
where $X \equiv \frac{1}{2} g^{\mu \nu} \partial_{\mu}\phi \partial_{\nu}\phi = -\dot{\phi}^2/2$ and $p(\phi,X)$ is a general function of $\phi$ and $X$. Note that the choice $p = -X -V(\phi)$ corresponds to the standard case with canonical kinetic term. The stress-energy tensor is defined as usual:
\begin{equation}
\label{eq_non_standard:Stress_energy_tensor}
T_{\mu \nu } \equiv -\frac{2}{\sqrt{|g|}} \frac{\delta \mathcal{S}_m}{\delta g^{\mu \nu}} \ ,  
\end{equation}
and allows to express the pressure and the energy density associated with the scalar field:	
	\begin{align}
	\label{eq_non_standard:p_phi}
		p&=p(\phi,X) \ ,\\
	\label{eq_non_standard:rho_phi}
		\rho(\phi,X)& =2 X p_{,X} - p= -\dot{\phi}^2 p_{,X}-p \ , 
	\end{align}
where $p_{,X} \equiv \partial p / \partial X$. The evolution of this system is completely specified by Einstein Equations. In the case of interest we obtain:
\begin{align}
\label{eq_non_standard:Friedmann}
	H^2&=\frac{\kappa^2}{3}\rho \ , \qquad\qquad -2\dot{H}=\kappa^2(\rho+p) = 2 \kappa^2 X p_{,X}\ . 
\end{align}
A phase of inflation corresponds to a nearly constant $H$ or equivalently a nearly negligible $\dot{H}$. This can be translated into a condition to achieve an inflationary stage by requiring $\epsilon_H \equiv -\dot{H}/H^2 \ll 1$. Using Eq.~\eqref{eq_non_standard:Friedmann} it becomes:
\begin{align}
\label{eq_non_standard:inflationcond}
	-\frac{\dot{H}}{H^2}&=\frac{3}{2}\frac{\rho + p}{ \rho} =\frac{2 X p_{,X} \kappa^2 }{2H^2} \ . 
\end{align}
Notice that inflation takes place when $\rho\simeq-p$. This condition can be either satisfied for negligible kinetic energy $\dot{\phi}^2 \simeq 0$ or in the limit $p_{,X} \simeq 0$. \\

More specifically in the case of a standard kinetic term, we have $p(\phi,X) = -X - V(\phi)$ which implies $p_{,X} = -1 $ and $\rho(\phi, X) = -X + V(\phi)$, an inflationary stage can only be realized for $\dot{\phi}^2  \ll  V(\phi)$. However, in the more general case, referred to as \emph{generalized kinetic term inflation} which is the focus of this paper, the condition of a normalized kinetic term, i.e. $p_{,X} = -1 $, is relaxed. As a consequence the condition for inflation $-\dot{H}/H^2\ll 1$ can also be satisfied with $p_{,X} \rightarrow 0$, and a whole new set of inflationary models, where inflation is actually driven by the kinetic term, can be defined. \\

Having described generalized kinetic term inflation, let us now discuss some fundamental difference with the standard case. A first difference appears in the expression of the speed of sound $c_s$ defined as $c_s^2\equiv p_{,X}/\rho_{,X}$. Using Eq.~\eqref{eq_non_standard:rho_phi} the speed of sound can be expressed as:
\begin{align}
\label{eq_non_standard:speedofsound}
	c_s^2=\frac{p_{,X}}{p_{,X}+2Xp_{,XX}}=\frac{1}{1+2X \partial \ln |p,_X|/\partial X} \; .
\end{align}
Obviously, for standard kinetic terms we have $c_s^2= 1$, this being no longer true for generalized kinetic term inflation. \\

Another main difference arises when considering the non-Gaussian features of primordial fluctuations\footnote{For details on the derivation of the three point correlation functions in single-field models see for example the classical papers of Maldacena~\cite{Maldacena:2002vr} and Weinberg~\cite{Weinberg:2005vy}, the well known work of Chen \textit{et. al}~\cite{Chen:2006nt} or the review of Chen~\cite{Chen:2010xka}.}. Interestingly, in the simplest realization of inflation the production of non-Gaussianities is highly suppressed~\cite{Maldacena:2002vr,Acquaviva:2002ud}. However, this result may change dramatically by considering models with non-standard kinetic terms, where sizable non-Gaussianities are allowed. \\

In order to discuss non-Gaussianities\footnote{The convention used in this paper is consistent with the notation used in~\cite{Chen:2010xka}.} we start by considering the three point correlation function of the curvature perturbation $\zeta_{\vec{k}}$ in momentum space:
\begin{equation}
	\label{eq_non_standard:bispectrum_def}
	\langle \zeta_{\vec{k}_1} \zeta_{\vec{k}_2} \zeta_{\vec{k}_3}\rangle \equiv B_{\zeta}(\vec{k}_1,\vec{k}_1,\vec{k}_1) \ ,
\end{equation}
where the function $B_{\zeta}(\vec{k}_1,\vec{k}_1,\vec{k}_1)$ is typically referred to as bispectrum. Using momentum conservation and rotational invariance the bispectrum can be expressed as:
\begin{equation}
 	B_{\zeta}(\vec{k}_1,\vec{k}_1,\vec{k}_1) = (2\pi)^3 \delta^{(3)}\left(\vec{k}_1+\vec{k}_2+\vec{k}_3\right) \, B_{\zeta} (k_1,k_2,k_3) \ .
 \end{equation} 
It is then customary to express $B(k_1,k_2,k_3)$ in terms of the dimensionless scalar power spectrum $\mathcal{P}_s(k)$ (see Eq.~\eqref{eq_formula:def_power_spectra}) evaluated at some fiducial scale $k_*$ as:
\begin{equation}
	B_{\zeta}(k_1,k_2,k_3) = (2\pi)^4 \mathcal{P}_s^2(k_*) \frac{S(k_1,k_2,k_3)}{(k_1 k_2 k_3)^2} \ ,
\end{equation}
where the dimensionless function $S(k_1,k_2,k_3)$ is the so-called ``shape function'' that depends on the three momenta appearing in the bispectrum. We can finally introduce the so-called ``nonlinearity parameter'' (introduced in~\cite{Babich:2004gb}) $f_{NL}$ defined as:
\begin{equation}
  \label{eq_non_standard:fnl_definition}
   S(K,K,K)\equiv \frac{9}{10} f_{NL} \ ,
\end{equation}
that corresponds to the amplitude of the bispectrum in momentum space in the equilateral momentum configuration\footnote{In this configuration the three-momenta satisfy $K = |k_1| = |k_2| = |k_3|$.}. Using this definition it is customary to extract $f_{NL}$ from $S(k_1,k_2,k_3)$ and to express the bispectrum as:
\begin{equation}
 	B_{\zeta}(k_1,k_2,k_3) = \frac{9}{10} f_{NL} \, (2\pi)^4 \mathcal{P}_s^2(k_*) \frac{S(k_1,k_2,k_3)}{(k_1 k_2 k_3)^2} \; ,
 \end{equation} 
where the shape function is now normalized to one in the equilateral configuration (\emph{i.e.} $S(K,K,K) = 1$). \\

It is possible to show~\cite{Chen:2006nt,Chen:2010xka} that for single field models, the leading contribution to $f_{NL}$ comes from equilateral configurations $f_{NL}^\text{equil}$. In particular the two dominant contributions (which are called $f_{NL}^{\lambda}$ and $f_{NL}^c$) can be expressed as:
\begin{equation}
  \label{eq_non_standard:leading_fnl}
  f_{NL}^{\lambda} = \frac{5}{81} \left( \frac{1}{c_s^2} - 1 -\frac{2 \lambda}{\Sigma}\right) \ , \qquad f_{NL}^{c} = -\frac{35}{108} \left( \frac{1}{c_s^2} - 1 \right) \ ,
\end{equation}
where $c_s^2$ is the speed of sound and where $\Sigma$ and $\lambda$ are defined as~\cite{Chen:2006nt,Chen:2010xka}:
\begin{eqnarray}
	\label{eq_non_standard:sigma_def}
	\Sigma &\equiv& X p_{,X} + 2 X^2 p_{,XX} = \frac{\epsilon_{H} H^2}{c_s^2} \ , \\
	\label{eq_non_standard:lambda_def}
  \lambda &\equiv&  X^2 p_{,XX} + \frac{2}{3}  X^3 p_{,XXX} \ .
\end{eqnarray}
Notice that in the case of standard kinetic terms (\textit{i.e.} for $p_{,X} =-1$) we have $\lambda = 0$ and $c_s^2 =1$ so that both these contributions are equal to zero. Higher order contributions are at least linear in the slow-roll parameters~\cite{Chen:2006nt,Chen:2010xka}. Planck measurements~\cite{Ade:2015xua,Ade:2015lrj,Ade:2015ava} constrains $f_{NL}^\text{equil}$ to be $|f_{NL}^\text{equil}| < |-4 \pm 43|$ at 68$\%$ CL.\\

To conclude this review of generalized kinetic term inflation, let us present an explicit example, the Dirac-Born-Infeld (DBI) model~\cite{Silverstein:2003hf,Alishahiha:2004eh} belonging to the category of brane inflation motivated by string theory. More precisely, the inflationary expansion of the Universe is due to the motion of a D3-brane in a AdS-like throat described by the DBI Lagrangian in a type IIB string theory. DBI inflation is specified by:
\begin{equation}
\label{eq_non_standard:pDBI}
p(X,\phi) = \frac{1}{f(\phi)} \left[ 1 - \sqrt{1+2f(\phi) X}\right] - V(\phi) \ ,
\end{equation}
where $f(\phi)=\lambda/\phi^4$ and $V(\phi)$ is the potential. For this particular class of models $p_{,X}$ is thus given by:
\begin{equation}
 	p_{,X} = -\frac{1}{\sqrt{1+2fX}} \ .
\end{equation} 
The DBI model of inflation has a non-trivial speed of sound $c_s^2=1+2fX$ and allows the production of sizable non-Gaussianities.

\section{Generalized $\beta$-function formalism.\label{sec:formalism}}
Let us now describe generalized kinetic term inflation using the $\beta$-function formalism presented in~\cite{Binetruy:2014zya}. Using the Hamilton-Jacobi approach of Salopek and Bond~\cite{Salopek:1990jq}, the field $\phi$ is taken as the clock to describe the system: the function $\phi(t)$ is inverted to express time in terms of the field $\phi$. In this framework, the Hubble parameter is a function of the field and it is useful to rewrite by introducing the superpotential $W(\phi) \equiv -2H(\phi)$ so that Friedmann equation reads:
\begin{equation}
	 \label{eq_formalism:friedman}
	\frac{3}{4\kappa^2} W^2 = \rho \ . 
\end{equation}
Differentiating $W(\phi)$ and using equation~\eqref{eq_non_standard:Friedmann} leads to:
	\begin{equation}
		-p_{,X} \, \dot{\phi}=\frac{W_{,\phi}}{\kappa^2} \ . \label{eq_formalism:phi_dot}
	\end{equation}
In order to proceed with the formulation of the problem in terms of the $\beta$-function formalism we must solve\footnote{In general solutions will only exist locally \emph{i.e.} on time-spans during which $\phi$ evolves monotonically.} Eq.~\eqref{eq_formalism:phi_dot} for $\dot{\phi} $ \emph{i.e.} we have to express $\dot{\phi}$ as a function of $\phi$. Once this solution is found, we can proceed by expressing $p_{,X}(X,\phi)$ as a function of $\phi$ only.\\

In analogy with~\cite{Binetruy:2014zya}, and inspired by the generalization of the $\beta$-function formalism to models with non-minimal coupling with gravity discussed in~\cite{Pieroni:2015cma}, we define the $\beta$-function associated with our model as:
	\begin{equation}
		\beta(\phi)\equiv\kappa \left(-p_{,X}\right)^{1/2} \frac{\textrm{d} \phi}{\textrm{d} \ln a} = - \frac{2}{\kappa} \left(-p_{,X}\right)^{-1/2} \frac{W_{,\phi}}{W} \ . \label{eq_formalism:beta_phi}
	\end{equation}
Note that other definitions for the $\beta$-function exist in the literature (see for example~\cite{Garriga:2015tea}). Our specific choice is motivated by requirement of invariance of the $\beta$-function under a field reparameterization. This is explicitly showed in Appendix~\ref{sec:appendix_theory} along with more arguments to justify the definition of Eq.~\eqref{eq_formalism:beta_phi}. By substituting Eq.~\eqref{eq_non_standard:Friedmann} and Eq.~\eqref{eq_formalism:phi_dot} into Eq.~\eqref{eq_non_standard:rho_phi} we can express the Hamilton-Jacobi equation as:
	\begin{equation}
		\frac{3}{2}W^2(\phi) + \frac{2}{\kappa^2}\left(p_{,X}\right)^{-1}W^2_{,\phi}(\phi)=-2\kappa^2p \ ,\label{eq_formalism:kequpot}
 	\end{equation}
which has the same form of the Hamilton-Jacobi equation derived in~\cite{Binetruy:2014zya}. Moreover, it is interesting to notice that once again the equation of state has a compact expression in terms of the $\beta$-function. Indeed substituting Eq.~\eqref{eq_non_standard:rho_phi} and Eq.~\eqref{eq_formalism:phi_dot} into Eq.~\eqref{eq_non_standard:inflationcond} we get:
		\begin{equation} 
			\frac{\rho+p}{\rho}=\frac{4\kappa^2}{3}\left(-p_{,X}\right)^{-1} \frac{\dot{\phi}^2}{W^2(\phi)} =  \frac{\beta^2(\phi)}{3} \ .  \label{eq_formalism:master}
		\end{equation}
This equation shows that the beta function is indeed the function that controls inflation! In particular inflation is realized in the vicinity of a zero of this function. As a consequence, the discussion in terms of universality classes may be properly extended to this generalized case. A review of most of the useful quantities to describe inflation expressed in terms of the $\beta$-function formalism is presented in Appendix~\ref{sec:appendix_formula}.\\

At this point is important to stress that while the definition of Eq.~\eqref{eq_formalism:beta_phi} is similar in form to the one given in~\cite{Pieroni:2015cma}, these two equations are substantially different. Indeed in the treatment of~\cite{Pieroni:2015cma} we have only considered models where the function $\left(-p_{,X}\right)^{1/2}$ \emph{only depends on} $\phi$! Conversely in the present work we are interested in discussing models where $\left(-p_{,X}\right)^{1/2}$ depends both on $\phi$ and $\dot{\phi}$. As a consequence the problem cannot be directly reduced to the case of the simplest realization of inflation! \\

\section{Models with non-standard kinetic terms.\label{sec:general}}
In what follows, we show how the $\beta$-function formalism is applied to some concrete examples of inflationary models with non-standard kinetic terms. The main difference with respect to the standard case discussed in~\cite{Binetruy:2014zya} is that a universality class of inflationary model is not only fixed by $\beta(\phi)$ but we also need to specify a parametrization for $p_{,X}(X,\phi)$. As explained in Sec.~\ref{sec:formalism}, once this parametrization is fixed we can proceed by solving Eq.~\eqref{eq_formalism:phi_dot} for $\dot{\phi}$ (or equivalently for $X$) and the whole dynamics can be expressed as a function of $\phi$ only. \\

It is important to stress that by fixing some parametrization for the $\beta$ function we specify a set of universality classes for inflationary models. However, as we also need to specify an explicit expression for $p_{,X}$, this universality only holds among models that share a similar expression for the non-standard kinetic term\footnote{Different parameterizations for $p_{,X}(X,\phi)$ lead to different $c_s$ that for example lead to a different amplitude of the scalar power spectrum (see Eq.~\eqref{eq_formula:scalar_spectrum}) and to a different value for $n_s $ (see Eq.~\eqref{eq_formula:ns}).}. In this sense the classification proposed in this paper is slightly different from the one proposed in~\cite{Binetruy:2014zya}.\\

Let us consider a quite general parametrization for $p_{,X}$:
\begin{equation}
	\label{eq_general:parametrization}
	p_{,X} = g(\phi) \left[ 1+2 f(\phi) X \right]^{\alpha}  \ .
\end{equation}
This parametrization has been chosen to recover two well-known examples of models of inflation with non-standard kinetic terms:
\begin{itemize}
 	\item By fixing $\alpha = -1/2$, $f=\lambda/\phi^4$ and $g(\phi)=-1$ we recover the DBI model~\cite{Silverstein:2003hf,Alishahiha:2004eh}.
 	\item By fixing $\alpha = -1/2$ and $f(\phi)=1$ and $g(\phi)=-V(\phi)$ we recover the case of tachyonic inflation~\cite{Gibbons:2002md,Padmanabhan:2002cp,Steer:2003yu}.
 \end{itemize} 
As a first step we can thus write Eq.~\eqref{eq_formalism:phi_dot} for the models described by the parametrization of Eq.~\eqref{eq_general:parametrization} (in the following we fix $\kappa^2=1$):
\begin{equation}
	\label{eq_general:phi_dot_param}
	-2X\left( 1 +2f(\phi)X\right)^{2\alpha} = \frac{W_{,\phi}^2}{g^2(\phi)} \ .
\end{equation}
As a consequence, once $\alpha$ is fixed we are able to solve (at least piecewise) Eq.~\eqref{eq_general:phi_dot_param} and we can thus express $X$ as a function of $\phi$. The inflationary model is then completely specified once an explicit parametrization for the $\beta$-function is known. Once the beta function is set, the superpotential can be computed by solving:
\begin{equation}
	\label{eq_general:beta}	
		-\frac{\kappa}{2}\beta(\phi) \left[- g(\phi)\right]^{1/2}  =  \left[ 1+2 f(\phi) X \right]^{-\alpha/2} \frac{W_{,\phi}}{W} \ .
\end{equation}
The speed of sound is then given by:
\begin{equation}
	\label{eq_general:speed_of_sound}
	c_s^2 = \frac{1 + 2fX}{1+2fX(1+2\alpha)} \ ,
\end{equation}
and it is also possible to show that:
\begin{equation}
	\label{eq_general:non_gaussianities}
	\frac{2\lambda}{\Sigma} = \frac{4 \alpha c_s^2 f X }{(1+2fX)^2} \left( 1 + \frac{4}{3} fX(\alpha + \frac{1}{2}) \right) \ , 
\end{equation}
where $\lambda $ and $\Sigma$ are defined respectively in Eq.~\eqref{eq_non_standard:sigma_def} and in Eq.~\eqref{eq_non_standard:lambda_def}.\\

As anticipated at the beginning of this section, universality classes of inflationary models can be completely specified by fixing $p_{X}$ and $\beta(\phi)$. It the rest of this section we first focus on a specific choice of parameter ($\alpha = -1/2$, $f=\lambda/\phi^4$ and $g(\phi)=-1$) that corresponds to ``DBI-like models''. In particular we show that in terms of the $\beta$-function formalism we can easily recover the analysis of~\cite{Silverstein:2003hf,Alishahiha:2004eh} and we explain how this analysis can be generalized. A similar treatment for models with $\alpha = -1/2$, $f=1$ and $g(\phi)=-V(\phi)$, \emph{i.e.} ``Tachyonic-like models'', is carried out in Appendix~\ref{sec:appendix_tachyon}. After this discussion, we consider a different class of models that can again be obtained from the parametrization of Eq.~\eqref{eq_general:parametrization}. In particular, we show that it is possible to define a set of inflationary models with $c_s \neq 1$ and small non-Gaussianities.

\subsection{DBI-like models.\label{sec:DBI}}
Let us consider the parametrization of Eq.~\eqref{eq_general:parametrization} and  fix $\alpha=-1/2$, $f(\phi)=\lambda/\phi^4$, $g(\phi)=-1$. In this case Eq.~\eqref{eq_general:phi_dot_param} reduces to:
	\begin{equation}
		-2X=W_{,\phi}^2\left( 1 + 2fX\right) \ ,
	\end{equation}
which allows to express $X$ as a function of $\phi$ only:
	\begin{equation}
		\label{eq_DBI:x_of_phi}
		-2X = \frac{W_{,\phi}^2}{1 + \lambda W_{,\phi}^2/\phi^4} \ .
	\end{equation}
Substituting this equation into the expression for the $\beta$-function given in Eq.~\eqref{eq_general:beta} we get:
\begin{equation}
	\label{eq_DBI:beta}	
		-\frac{\beta(\phi)}{2}  = \frac{W_{,\phi}}{W \left(1 + \lambda W_{,\phi}^2/\phi^4 \right)^{1/4}} \ .
\end{equation}
This equation can be inverted to express $W_{,\phi}$ as a function of $\beta$ and $W$:
\begin{equation}
	\label{eq_DBI:beta_super}
	W_{,\phi} = - \sqrt{\frac{\lambda}{2}} \frac{ W^2 \beta^2}{ 4 \phi^2} \left(1 + \sqrt{1 + \frac{64 \, \phi^8 }{\lambda^2 \beta^4 W^4} } \  \right)^{1/2} \ .
\end{equation}
As a consequence, once the parametrization of the $\beta$-function is fixed, we can solve this differential equation to obtain the superpotential.\\

Similarly we can substitute Eq.~\eqref{eq_DBI:x_of_phi} into Eq.~\eqref{eq_general:speed_of_sound} and into Eq.~\eqref{eq_general:non_gaussianities} to get respectively:
\begin{eqnarray}
	\label{eq_DBI:speed_of_sound}
	c_s^2 &=& \frac{1}{1 + \lambda W_{,\phi}^2/\phi^4} \ , \\
	\label{eq_DBI:ng}
	\frac{2\lambda}{\Sigma} &=& \frac{\lambda W_{,\phi}^2/\phi^4}{1 + \lambda W_{,\phi}^2/\phi^4} \ .
\end{eqnarray}
The expressions for all the other relevant quantities to describe inflation can be derived using the general formula that are reported in Appendix~\ref{sec:appendix_formula}. In particular, comparing Eq.~\eqref{eq_formula:pressure}	with Eq.~\eqref{eq_non_standard:pDBI} it is easy to show that the potential can be expressed as:
\begin{equation}
	\label{eq_DBI:potential}
	V(\phi) = \frac{\phi^4}{\lambda} \left[1 - \frac{1}{\sqrt{1 + \lambda W_{,\phi}^2/\phi^4} } \right] + \frac{3}{4} W^2(\phi) \left[ 1 - \frac{\beta^2}{3} \right]\;.
\end{equation}

At this point we can finally define a set universality class of models by specifying some explicit parameterizations of the $\beta$-function. For example we can assume that in a neighborhood of the fixed point the $\beta$-function can be expressed as:
\begin{align}
\label{eq_DBI:betaexpansion}
\beta(\phi) &\simeq a \phi^n  \ .
\end{align}
The asymptotic expressions for the superpotential can be computed using Eq.~\eqref{eq_DBI:beta_super}. In particular we can distinguish three different regimes (that, as we show in the following, correspond to different values of $n$):
\begin{itemize}
 	\item For $\phi^2/(\beta W) \ll 1$ Eq.~\eqref{eq_DBI:beta_super} can be approximated as:
 		\begin{equation}
			\label{eq_DBI:superpot_prime_small}
			W_{,\phi} \simeq - \sqrt{\lambda} \frac{ W^2 \beta^2}{ 4 \phi^2}  \ .
		\end{equation}
		We can thus integrate this equation to get:
 		\begin{align}
			\label{eq_DBI:superpot_small_1/2}
			W(\phi) &\simeq W_{\textrm{f}} \left[ 1 + \frac{W_{\textrm{f}} \, a^2 \,\sqrt{\lambda}}{4} \ln(\phi) \right]^{-1} \ , \qquad \text{for } \qquad n = \frac{1}{2}  \ , \\ 
			\label{eq_DBI:superpot_small}
			W(\phi) &\simeq \left[ \frac{1}{W_{\textrm{f}}} + \frac{a^2 \,\sqrt{\lambda}}{4 (2n -1)} \phi^{2n-1} \right]^{-1} \ , \qquad \text{for } \qquad n\neq \frac{1}{2} \ . 
		\end{align}
		Notice that Eq.~\eqref{eq_DBI:superpot_small} gives two asymptotic behaviors for the superpotential:
		\begin{align}
			\label{eq_DBI:superpot_fin_1}
			W(\phi) &\simeq \frac{4(2n-1)}{a^2 \sqrt{\lambda}} \phi^{1-2n} \ , \qquad \text{for } \qquad 0 \leq n < \frac{1}{2}  \ , \\ 
			\label{eq_DBI:superpot_fin_2}
			W(\phi) &\simeq W_{\textrm{f}} \left[ 1 - \frac{a^2 \,\sqrt{\lambda}  W_{\textrm{f}} }{4 (2n -1)} \phi^{2n-1} \right] \ , \qquad \text{for } \qquad n > \frac{1}{2} \ . 
		\end{align}
		Comparing Eq.~\eqref{eq_DBI:superpot_fin_2} with the condition $\phi^2/(\beta W) \ll 1$ it should be clear that this expression only holds for $n<2$! Notice also that the case $n = 0$ is special as in this case $\beta$ approaches a constant non-zero value. As discussed in~\cite{Binetruy:2014zya}, this case corresponds to power-law inflation~\cite{Lucchin:1984yf}. Finally we can use Eq.~\eqref{eq_DBI:potential} to compute the potential associated to the different expressions of $W$:
		\begin{align}
			\label{eq_DBI:pot_fin_small_0}
			V(\phi) &\simeq \frac{12}{\lambda a^4 } \left( 1- \frac{a^2}{3}\right) \phi^{2} \ , \qquad \text{for } \qquad n = 0 \ , \\ 
			\label{eq_DBI:pot_fin_small_1}
			V(\phi) &\simeq \frac{12(1 - 2n)^2}{\lambda a^4 } \phi^{2(1-2n)} \ , \qquad \text{for } \qquad 0 \leq n < \frac{1}{2}  \ , \\ 
			\label{eq_DBI:pot_fin_small_1/2}
			V(\phi) &\simeq \frac{3 W_{\textrm{f}}^2}{4} \left[1 + \frac{W_{\textrm{f}} \, a^2 \, \sqrt{\lambda} }{4} \ln(\phi) \right]^{-2} \ , \qquad \text{for } \qquad n = \frac{1}{2} \ , \\ 
			\label{eq_DBI:pot_fin_small_2}
			V(\phi) &\simeq \frac{3 W_{\textrm{f}}^2}{4} \left[1 - \frac{W_{\textrm{f}} \, a^2 \, \sqrt{\lambda} }{2(2n-1)} \phi^{2n-1} \right] \ , \qquad \text{for } \qquad \frac{1}{2} \leq n < 2 \ .
		\end{align}
		These cases correspond to some of the ansatzs considered in~\cite{Silverstein:2003hf}. In particular $n = 0 $ corresponds to the case B, $0 \leq n < 1/2$ is a generalization of case B, $1/2 \leq n < 2$ is a generalization of case A and $n=1/2$ corresponds to a case that was not considered in~\cite{Silverstein:2003hf}.\\

		It is also interesting to point out that for all these cases $c_s^2$ goes to zero during inflation. As a consequence all these models predict the generation of large non-Gaussianities.
	\item For $\phi^2/(\beta W) \gg 1$ Eq.~\eqref{eq_DBI:beta_super} can be approximated as:
 		\begin{equation}
			\label{eq_DBI:superpot_prime_large}
			W_{,\phi} \simeq - \frac{ W \beta}{ 2}  \ .
		\end{equation}
		We can thus integrate this equation to get:
 		\begin{equation}
			\label{eq_DBI:superpot_large}
			W(\phi) \simeq W_{\textrm{f}} \exp\left[ -\frac{a }{2} \frac{\phi^{n+1}}{n+1} \right] \ . 
		\end{equation}
		Comparing Eq.~\eqref{eq_DBI:superpot_large} with the condition $\phi^2/(\beta W) \gg 1$ it should be clear that this expression only holds for $n>2$! We can then use Eq.~\eqref{eq_DBI:potential} to compute the asymptotic expression for the potentials associated with these models:
		\begin{align}
			\label{eq_DBI:pot_fin_large}
			V(\phi) &\simeq \frac{3 W_{\textrm{f}}^2 }{4 }\exp\left[ - a  \frac{\phi^{n+1}}{n+1} \right] \ .
		\end{align}	
		Expanding the potential at the first order we find that this case corresponds to a generalization of case A of~\cite{Silverstein:2003hf}. For the models of this class $c_s^2$ approaches $1$ and using Eq.~\eqref{eq_DBI:ng} it is easy to show that these models predict negligible non-Gaussianities. 
	\item Finally we can consider the case $\phi^2/(\beta W) \simeq C$ where $C$ is a constant of order $1$. This clearly corresponds to $n=2$, and thus Eq.~\eqref{eq_DBI:beta_super} reads: 
 		\begin{equation}
			\label{eq_DBI:superpot_prime_med}
			W_{,\phi} \simeq - D \sqrt{\lambda} \frac{ W^2 \beta^2}{ \phi^2}  \ ,
		\end{equation}
		where we have defined $D \equiv [(1 + \sqrt{1 + C})/8]^{1/2}$. We can thus integrate this equation to get:
 		\begin{equation}
			\label{eq_DBI:superpot_med}
			W = W_{\textrm{f}} \left[ 1 - \frac{D a^2 \,\sqrt{\lambda}  W_{\textrm{f}} }{ 3 } \phi^{3} \right] \ .
		\end{equation}
		Once again we can use Eq.~\eqref{eq_DBI:potential} to compute the asymptotic expression for the potentials associated with these models:
		\begin{align}
			\label{eq_DBI:pot_fin_med}
			V(\phi) &\simeq \frac{3 W_{\textrm{f}}^2 }{4 } \left[ 1 - \frac{2 D a^2 \,\sqrt{\lambda}  W_{\textrm{f}} }{ 3 } \phi^{3} \right] \ .
		\end{align}	
		Again these models can be seen as a generalization of case A of~\cite{Silverstein:2003hf}. \\

		As for the models of this class $\lambda W_{,\phi}^2/\phi^4 = \lambda^2 D^2 W_{\textrm{f}}^4 a^4$ we have 
		\begin{equation}
			c_s^2 = \frac{1}{1 + \lambda^2 D^2 W_{\textrm{f}}^4 a^4} \ .
		\end{equation}
		As a consequence $c_s^2$ is different from $1$ and is finite. Moreover, we can use Eq.~\eqref{eq_DBI:ng} and Eq.~\eqref{eq_non_standard:leading_fnl} to show that for these models we have:
		\begin{equation}
		 	f_{NL}^{\lambda} = \frac{5}{81} \left( \frac{\lambda^4 D^4 W_{\textrm{f}}^8 a^8 }{1 +  \lambda^2 D^2 W_{\textrm{f}}^4 a^4} \right) \ , \qquad f_{NL}^{c} = \frac{35}{108} \left( \lambda^4 D^4 W_{\textrm{f}}^8 a^8 \right) \ .
		 \end{equation} 
		These models are thus predicting non-zero non-Gaussianities which are not arbitrarily large. 
 \end{itemize}  
Notice that all the models considered in this section are obtained assuming that in the neighborhood of the fixed point the $\beta$-function can be parameterized as in Eq.~\eqref{eq_DBI:betaexpansion}. As a consequence new classes of models can be defined if we consider different parameterizations.

\subsection{A new class of models.\label{sec:new_models}}
While the models of Sec.~\ref{sec:DBI} have been useful in order to explain the application of the formalism, it is fair to point out that they typically predict~\cite{Li:2013cem} values of cosmological observables (in particular $n_s$ and $r$ and/or a huge amount of non-Gaussianities) that are not in good agreement with Planck constraints~\cite{Ade:2015xua,Ade:2015lrj}. In order to define a class of models that satisfies this requirements we can consider some other models that can be described by the parametrization of Eq.~\eqref{eq_general:parametrization}. \\

For this purpose we fix $g(\phi) = \gamma/\phi$ and we us assume that once the dynamics is expressed in terms of $\phi$, in a neighborhood of the fixed point and up to subleading corrections we have $2fX \simeq - C^2$ where $C$ is a constant. In the following we will check the consistency of this assumption. \\

We proceed by using Eq.~\eqref{eq_general:phi_dot_param} to get: 
\begin{equation}
	\label{eq_new:X_superpot}
	-2X = \frac{ W_{\phi}^2}{g^2(\phi)} (1 + 2fX)^{- 2 \alpha}\simeq \phi^2 \frac{ W_{\phi}^2}{\gamma^2} (1 -C^2 )^{- 2 \alpha} \ .
\end{equation}
The relation between the $\beta$-function and the potential is then given by Eq.~\eqref{eq_general:beta} and it reads:
\begin{equation}
	\label{eq_new:beta_superpot}
  \beta(\phi) = - 2 (- p_{,X})^{1/2} \frac{W_{,\phi}}{W} \simeq - 2 \sqrt{\frac{\gamma}{\phi}} (1 - C^2)^{\alpha/2} \frac{W_{\phi}}{W} \ .
\end{equation}
Using the definition of $c_s^2$ and $\frac{2 \lambda}{\Sigma}$ it is easy to show that we have:
\begin{equation}
	\label{eq_new:cs_fnl}
  c_s^2 \simeq \frac{1 - C^2}{1 - (2\alpha +1) C^2} \ , \qquad  \frac{2 \lambda}{\Sigma} \simeq -\frac{2}{3}  \frac{(4 \alpha - 1) C^2}{1 - (2\alpha +1) C^2} \ .
\end{equation}
Note that we have still not fixed any particular parameterization for $f(\phi)$ but we only required that during the evolution $X(\phi)$ approaches $-C^2/(2 f)$. The explicit parameterizations of $f(\phi)$ and $X(\phi)$ are thus fixed a posteriori when we fix the expression for $\beta(\phi)$. The value of $\alpha$ has not been specified either. For example it is possible to choose $\alpha = 1/4$ to set $2 \lambda/ \Sigma = 0$. As a consequence, the production of non-Gaussianities in these models is only controlled by $c_s^2$. \\

At this point, in order to specify a class of inflationary models and to compute the associated predictions for cosmological observables, a parametrization for $\beta(\phi)$ has to be chosen. In the following we focus on the case $\beta = - \gamma/\phi$. Let us start by computing $N$, number of e-foldings (using Eq.~\eqref{eq_formula:Nefold}):
\begin{equation}
  N = - \int_{\phi_\textrm{f}}^\phi (-p_{,X})^{1/2} \frac{\textrm{d} \hat{\phi}}{\beta(\hat{\phi})} = (1 - C^2)^{\alpha/2} (\phi - \phi_\textrm{f}) \ .
\end{equation}
Similarly we can use Eq.~\eqref{eq_new:beta_superpot} to get the explicit expression for $W(\phi)$:
\begin{equation}
  W = W_{\textrm{f}} \exp \left\{ - \frac{\gamma^2 (1 - C^2)^{\alpha/2} }{2} \left( \frac{1}{\phi} -\frac{1}{\phi_{\textrm{f}}} \right) \right\} \ .
\end{equation}
The lowest order expression for $W_{, \phi} $ is given by:
\begin{equation}
	W_{, \phi} \simeq - \frac{\gamma^2 (1 -C^2)^{\alpha/2}}{2 \phi^2} \ ,
\end{equation}
and  $X(\phi)$ can also be directly computed to the lowest order using Eq.~\eqref{eq_new:X_superpot}:
\begin{equation}
	-2X \simeq \frac{\gamma^2}{\phi^2 \, (1 -C^2)^{\alpha}}  \ .
\end{equation}
In order to be consistent with the assumption $2fX \simeq −C^2$ we have:
\begin{equation}
	f(\phi) \simeq \frac{\phi^2 \, (1 -C^2)^{\alpha}}{\gamma^2}\, C^2 \ .
\end{equation}
Finally we can compute the lowest order expressions for $n_s$ and $r$ using Eq.~\eqref{eq_formula:ns} and Eq.~\eqref{eq_formula:ns} respectively:
\begin{equation}
  n_s -1 \simeq -\frac{2}{N^2} - \frac{\gamma^2 (1 - C^2)^{\alpha/2}}{N^2} \ , \qquad r \simeq 8 c_s \beta^2 \simeq \frac{\mathcal{O}(1)}{N^2} \ .
\end{equation}
For the typical value of $N$ (\emph{i.e.} $N \simeq 60 $) we can find a large region of the parameter space that gives values of $n_s$ and $r$ that are in good agreement with Planck constraints~\cite{Ade:2015xua,Ade:2015lrj}. Moreover, as shown in Eq.~\eqref{eq_new:cs_fnl}, these models are also predicting $c_s^2 \neq 1$ and sizable but not huge non-Gaussianities.

\section{Conclusions.\label{sec:conclusions}}
Models with non-standard kinetic term may predict both a deviation from $c_s^2 =1$ and the production of sizable non-Gaussianities. As these quantities can be directly constrained using data~\cite{Ade:2015xua,Ade:2015lrj,Ade:2015ava}, it is interesting to understand whether these models can still be considered as viable or if they are completely ruled out. The $\beta$-function formalism was originally introduced in~\cite{Binetruy:2014zya} to describe single field inflationary models where the inflaton has a canonical kinetic term and a minimal coupling with gravity\footnote{The extension to non-minimal coupling is discussed in~\cite{Pieroni:2015cma}.}. In this work we have proposed an extension of this formalism in order to include inflationary models with non-standard kinetic terms. In particular we have shown that the description of inflation in terms of RG equation still holds, allowing for the definition of universality classes among different theoretical realizations of inflation.\\

In this formalism, any model is completely defined by the form of its non-standard kinetic term (which is parameterized by $p_{,X}$) and by the $\beta$-function. As a first step we have explicitly shown that some well-known models, such DBI (and tachyonic) inflation can be easily reproduced and generalized in terms of this formalism. Moreover, as our approach provides a powerful bottom-up method to construct new models, we have shown (using the parametrization of Eq.~\eqref{eq_general:parametrization}) that a large class of new models can be introduced. We have also shown that is possible to define models with $c_s^2 \neq 1$ and with non-Gaussianities that are non-trivial but small enough to potentially agree with data. The definition of a concrete (and hopefully theoretically well-motivated) inflationary model arising from some high energy theory that belongs to this class offers an interesting possibility for future works. \\

A further generalization would be to extend our formalism to multi-fields models of inflation\footnote{The analysis of~\cite{Bourdier:2013axa} and~\cite{Garriga:2015tea} can provide a useful guideline to achieve this result.}. It would also be interesting to discuss the extension of the formalism both to models with a non-minimal \emph{kinetic} coupling between the inflaton and gravity (as in the case of ``new Higgs'' inflation~\cite{Germani:2010gm,Germani:2010ux}), and to models of modified gravity. These analysis are left to future works.

\subsection*{ Acknowledgements}
M.P. would like to thank E. Kiritsis and F. Vernizzi for useful discussions.\\
P.B. and M.P. acknowledge the financial support of the UnivEarthS
Labex program at Sorbonne Paris Cit\'e (ANR-10-LABX-0023 and
ANR-11-IDEX-0005-02). J.M. is supported by Principal’s
Career Development Scholarship and Edinburgh Global Research Scholarship. \\
J.M. and M.P. want to dedicate this paper to the memory of Pierre Bin\'etruy who passed away right after the end of the reviewing process. Pierre was not only a great physicist but also a great person and probably the best ``boss'' we could have dreamed of.

\newpage
\appendix

\section{Useful formula.\label{sec:appendix_formula}}
In this section we present a summary of the main results for a general model. Predictions are parametrized in terms of the function $\beta(\phi)$ and of $p_{,X}$. As explained in Sec.~\ref{sec:formalism} and as explicitly shown in Sec.~\ref{sec:general}, once an explicit expression for $p_{,X}$ is fixed, we can express $\dot{\phi}$ as a function of $\phi$. As consequence everything can be expressed as an explicit function of $\phi$.
\begin{itemize}
	\item The superpotential is obtained by solving the equation: 
	  	\begin{equation}
	  	\label{eq_formula:beta}	
			\beta(\phi) =-\frac{2}{\kappa}\left(-p_{,X}\right)^{-1/2}\frac{W_{,\phi}}{W} \;.		\end{equation}	
	\item The pressure $p(\phi)$ is: 
	  	\begin{equation}	
	  	\label{eq_formula:pressure}	
       			p(\phi) =-\frac{3}{4\kappa^{2}}\left(1-\frac{\beta^{2}(\phi)}{3}\right)W^{2}.
 		\end{equation}
	\item The number of e-foldings is: 
	  	\begin{equation}	
       			N =-\kappa\int_{\phi_{f}}^{\phi}\left(-p_{,X}\right)^{1/2}\frac{\mathrm{d}\phi'}{\beta(\phi')} .\label{eq_formula:Nefold}
 		\end{equation}
\end{itemize}
We then express the dimensionless power spectra\footnote{The dimensionless scalar and tensor power spectra at horizon crossing are respectively defined as:
\begin{equation}
	\label{eq_formula:def_power_spectra}
 	\mathcal{P}_s=\frac{\kappa^2 H^2}{8\pi^2 \epsilon_H c_s } \ , \qquad \qquad \mathcal{P}_t=\frac{2 \kappa^2 H^2}{\pi^2} \ .
 \end{equation}
 } and their spectral indexes as:
\begin{itemize}
	\item Scalar power spectrum: 
	  	\begin{equation}	
	  			\label{eq_formula:scalar_spectrum}
       			\mathcal{P}_s=\frac{\kappa^2}{16\pi^2}W^2 \frac{1}{c_s\beta^2(\phi)}.
 		\end{equation}
	\item Scalar spectral index: 
	  	\begin{equation}	
	  			\label{eq_formula:ns}
       			n_{s}-1 \simeq \frac{\left[-\beta^{2}(\phi)-\frac{\beta(\phi)}{\kappa}\left(-p_{,X}\right)^{-1/2}\frac{\textrm{d}}{\textrm{d}\phi} \ln\left(c_s\beta^2(\phi)\right)\right]}{1-\frac{1}{2} \beta^2(\phi) -\frac{\beta(\phi)}{\kappa} \left(-p_{,X}\right)^{-1/2}\frac{c_s,_\phi}{c_s}}
.
 		\end{equation}	
	\item Tensor power spectrum: 
	  	\begin{equation}	
       			\mathcal{P}_t=\frac{\kappa^2}{2\pi^2}W^2(\phi)\;.
 		\end{equation}	
	\item Tensor spectral index: 
	  	\begin{equation}	
       			n_{t} \simeq -\frac{\beta^2(\phi)}{1-\frac{1}{2}\beta^2(\phi)-\frac{\beta(\phi)}{\kappa}\left(-p_{,X}\right)^{-1/2}\frac{c_s,_\phi}{c_s}}  \ .
 		\end{equation}	
	\item Tensor-to-scalar ratio: 
	  	\begin{equation}	
	  			\label{eq_formula:r}
       			\mathrm{r} =8c_s\beta^{2}(\phi).
 		\end{equation}
\end{itemize}
To conclude this section, we briefly discuss slow-roll parameters.  ``Horizon-flow" parameters as introduced in~\cite{Schwarz:2001vv} are particularly suitable for k-inflation\footnote{Notice that in the referred paper $N \equiv \ln(a/a_i)$ while we defined $N \equiv - \ln(a/a_f)$.}.  They are defined as:
	\begin{equation}	
       		\epsilon_0\equiv H_*/H,\qquad\qquad  \epsilon_{i+1}\equiv - \frac{\textrm{d} \ln |\epsilon_i|}{\textrm{d} N} = {\beta (\phi) \over \kappa} \left(-p_{,X}\right)^{-1/2} \frac{\textrm{d} \ln |\epsilon_i|}{\textrm{d} \phi}\,,
 	\end{equation}
where $H_*$ is the Hubble parameter at some chosen time $t_*$. We compute $\epsilon_1$ and $\epsilon_2$:
	\begin{align}
		\epsilon_1 &=\frac{1}{H}\frac{\textrm{d}H}{\textrm{d}N}=\frac{1}{2} \beta^2(\phi),\\
		\epsilon_2 &=\frac{2\left(-p_{,X}\right)^{-1/2}}{\kappa}\frac{\textrm{d} \beta}{\textrm{d}\phi},
	\end{align}
where we used $\textrm{d}N=-\frac{\kappa}{\beta(\phi)}\left(-p_{,X}\right)^{1/2} \textrm{d}\phi$. 
In addition, we define the ``speed-of-sound" parameter $\epsilon_{c_s}$:
	\begin{equation}
		\epsilon_{c_s} =\frac{\beta(\phi)}{\kappa} \left(-p_{,X}\right)^{-1/2} \frac{\textrm{d} \ln c_s}{\textrm{d}\phi}.
	\end{equation}
Depending on the model, $\epsilon_{c_s}$ may be written in terms of the other parameters $\epsilon_i$. The spectral indexes and the tensor to scalar ratio in terms of these parameters read:
	\begin{align}
		n_s-1 &=\frac{-2\epsilon_1 -\epsilon_2 -\epsilon_{c_s}}{1 - \epsilon_1 -\epsilon_{c_s}},\\
		n_t &=\frac{-2 \epsilon_1}{1 - \epsilon_1 - \epsilon_{c_s}},\\
		r &= 16c_s\epsilon_1.
	\end{align}

\section{$\beta$-function parametrization and field redefinition.\label{sec:appendix_theory}}
In this appendix we present some theoretical motivations (which are also supported by some examples) to justify the definition of the beta-function given in Eq.~\eqref{eq_formalism:beta_phi}. As discussed in~\cite{Pieroni:2015cma,Pieroni:2016gdg}, if $p_{,X}$ is a function of $\phi$ only, say $-F(\phi)$, it is possible to define a new field $\varphi$ as:
	\begin{equation}
		\label{eq_theory:redefinition}
		\left( \frac{\textrm{d} \varphi }{ \textrm{d} \phi}\right) \equiv (-p_{,X})^{1/2} = F^{1/2}(\phi) \ ,
	\end{equation}
that has a canonical kinetic term. As a consequence, the $\beta$-function that describes the cosmological evolution of $\varphi$  can be defined directly following the procedure carried out in~\cite{Binetruy:2014zya}. In particular we have:
	\begin{equation}
		\tilde{\beta}(\varphi) \equiv \beta(\phi(\varphi)) = \kappa  \frac{\textrm{d} \varphi}{\textrm{d} \ln a} \ ,
	\end{equation} 
so that the evolution of the inflationary Universe is described in terms of an equation that has exactly the form of a renormalization group equation. In this case the $\beta$-function that describes the evolution of $\phi$ is given by:
\begin{equation}
	\beta(\phi) = \tilde{\beta}(\varphi(\phi)) = \kappa  \frac{\textrm{d} \varphi }{ \textrm{d} \phi}  \frac{\textrm{d} \phi}{\textrm{d} \ln a} = \kappa F^{1/2} (\phi) \frac{\textrm{d} \phi}{\textrm{d} \ln a} \ ,
\end{equation}
and therefore, we can equivalently work with either $\beta(\phi)$ or $\beta(\varphi)$. The definition of $\beta$ given in Eq.~\eqref{eq_formalism:beta_phi} is thus motivated by the request for this property. However, it is fair to mention that different definitions for $\beta$ (that indeed are perfectly valid) exist in the literature (see for example~\cite{Garriga:2015tea}).\\

As explained in Sec.~\ref{sec:non-standard} and in Sec.~\ref{sec:formalism}, in the case of a generalized kinetic term, $p_{,X}$ may be a general function of $X$ and $\phi$. As a consequence, in general we cannot perform the field redefinition of Eq.~\eqref{eq_theory:redefinition} and we cannot describe the dynamics in terms of a single scalar field with a canonical kinetic term. However, as discussed in Sec.~\ref{sec:non-standard} and in Sec.~\ref{sec:formalism}, once we have solved\footnote{The solution of this equation exists at least locally.} Eq.~\eqref{eq_formalism:phi_dot} for $\dot{\phi}$, it is still possible to express $\dot{\phi}$ as a function of $\phi$ only. After this procedure, even $p_{,X}$ can be expressed as a function of $\phi$ only and thus, using Eq.~\eqref{eq_theory:redefinition} we can define\footnote{While it is not always possible to find an analytical expression for $\phi(\varphi)$, in following we show that in some cases this expression exists.} a new field $\varphi$ whose cosmological evolution is described by:
\begin{equation}
 	\tilde{\beta}(\varphi) \equiv \beta(\phi(\varphi)) = \kappa  \frac{\textrm{d} \varphi}{\textrm{d} \ln a} \ ,
 \end{equation}
 that once again has exactly the form of a renormalization group equation. Notice that expressing Eq.~\eqref{eq_formalism:phi_dot} in terms of $\varphi$ we recover the usual expression for $\dot{\varphi}$ in terms of the superpotential
	\begin{equation}
			 \dot{\varphi}=\frac{\tilde{W}_{,\varphi}(\varphi)}{\kappa^2} \ . \label{eq_formalism:varphi_dot}
	\end{equation}
where $\tilde{W}(\varphi)\equiv W(\phi(\varphi))$. Using this equation we can directly get:
\begin{equation}
		\tilde{\beta}(\varphi) = \kappa  \frac{\textrm{d} \varphi}{\textrm{d} \ln a}  = -\frac{2}{\kappa} \frac{\tilde{W}_{,\varphi}}{\tilde{W}} \ ,
	\end{equation} 
and the usual Hamilton-Jacobi equation of~\cite{Binetruy:2014zya}:
	\begin{equation}
		-2\kappa^2\tilde{p}(\varphi) = \frac{3}{2}\tilde{W}^2(\varphi) -\frac{2}{\kappa^2}\tilde{W}^2_{,\varphi}(\varphi)= \frac{3}{2}\tilde{W}^2(\varphi) \left[ 1 - \frac{\tilde{\beta}^2(\varphi)}{3} \right] \ .\label{eq_formalism:varphi_HJ}\;.
 	\end{equation} 
At this point it is crucial to stress that even if it is possible to recover the expressions derived in~\cite{Binetruy:2014zya}, in general the field $\varphi$ has a non-standard kinetic term\footnote{An explicit example of this feature is shown at the end of this Appendix (see Eq.~\eqref{eq_non_standard_varphi_p}).}! In particular, while 
\begin{equation}
 (-\tilde{p}_{,X}(\varphi) )^{1/2} \left(\frac{\textrm{d} \phi}{\textrm{d} \varphi} \right) = 1 \ ,
\end{equation}
we still have $\tilde{p}_{,X}(X,\varphi) \neq -1$! As a consequence both $\tilde{p}_{,XX}(\varphi)\equiv p_{,XX}(\phi(\varphi))$ and $\tilde{p}_{,XXX}(\varphi)\equiv p_{,XXX}(\phi(\varphi))$ are generic functions of $\varphi$ and thus the model is still predicting $c_s^2 \neq 1$ and non-zero non-Gaussianities.\\

Let us clarify this aspect by considering some examples where an analytical expression for $\varphi$ can be computed. In particular, let us consider some of the cases of DBI-like inflation models presented in Sec.~\ref{sec:DBI}. We can start by using Eq.~\eqref{eq_DBI:x_of_phi} to express $p_{,X}$ as:
\begin{equation}
	-p_{,X} = \left( 1 + \lambda W_{,\phi}^2 /\phi^4 \right)^{-1} \ ,
\end{equation}
and the expression for the $\beta$-function reads:
\begin{equation}
\beta(\phi) = (- p_{,X})^{1/2} \frac{\textrm{d} \phi}{\textrm{d} \ln a} = \left( 1 + \lambda W_{,\phi}^2 /\phi^4 \right)^{-1/2}	\frac{\textrm{d} \phi}{\textrm{d} \ln a} \ .
\end{equation}
In order to cast the $\beta$-function in the canonic form we can proceed with the field redefinition of Eq.\eqref{eq_theory:redefinition} \emph{i.e.}:
\begin{equation}
	\label{eq:varphi}
	\textrm{d} \varphi = \frac{{\textrm{d} \phi}}{\sqrt{1 + \lambda W_{,\phi}^2 /\phi^4}}
\end{equation}
At this point we can set $\beta(\phi) \simeq a \phi^n$ and we show that it is possible to define some cases that admit an analytical expression for $\phi(\varphi)$: 
\begin{itemize}
\item As a first example we consider the case with $n=2$. In this case $W(\phi)$ is given by Eq.~\eqref{eq_DBI:superpot_med} \emph{i.e.}:
\begin{equation}
	W = W_{\textrm{f}} \left[ 1 - \frac{D a^2 \,\sqrt{\lambda}  W_{\textrm{f}} }{ 3 } \phi^{3} \right] \ ,
\end{equation}
and thus:
\begin{equation}
	W_{,\phi} \simeq - D a^2 \,\sqrt{\lambda}  W_{\textrm{f}}^2  \phi^{2}   \ .
\end{equation}
By substituting into Eq.~\eqref{eq:varphi} to get:
\begin{equation}
	\textrm{d} \varphi \simeq \frac{\textrm{d} \phi}{\sqrt{1 + \lambda^2 D^2 a^4 \, W_{\textrm{f}}^4 }} \, \qquad \longrightarrow \qquad  \varphi \simeq \frac{ \phi}{\sqrt{1 + \lambda^2 D^2 a^4 \, W_{\textrm{f}}^4 }} \ .
\end{equation}
Notice also that in this case $p_{,X}$ becomes nearly constant approaching the fixed point. 

\item In order to define two more examples, we start by setting $n > 2$. In this case $W(\phi)$ is given by Eq.~\eqref{eq_DBI:superpot_large} \emph{i.e.}:
\begin{equation}
	W  \simeq W_{\textrm{f}} \exp\left[ -\frac{a }{2} \frac{\phi^{n+1}}{n+1} \right] \simeq W_{\textrm{f}} - W_{\textrm{f}} \frac{a}{2} \frac{\phi^{n+1}}{n+1} \ ,
\end{equation}
and thus:
\begin{equation}
	W_{,\phi} \simeq - \frac{a \, W_{\textrm{f}}}{2} \phi^{n}   \ .
\end{equation}
We can thus substitute into Eq.~\eqref{eq:varphi} to get:
\begin{equation}
	\textrm{d} \varphi \simeq \frac{\textrm{d} \phi}{\sqrt{1 + \lambda a^2 W_{\textrm{f}}^2 \phi^{2n - 4} / 4 }} \ .
\end{equation}
At this point we can notice that setting $n = 5/2$ we have:
\begin{equation}
	\label{varphi5/2}
	\textrm{d} \varphi \simeq \frac{\textrm{d} \phi}{\sqrt{1 + \lambda a^2 W_{\textrm{f}}^2 \phi / 4 }} \ , \qquad \longrightarrow \qquad \varphi -\varphi_{\textrm{f}} = \frac{2}{C} \, \sqrt{1 + C \phi }  \ ,
\end{equation}
where $C \equiv \lambda a^2 W_{\textrm{f}}^2 / 4 $. Clearly this equation can be inverted to express $\phi$ as a function of $\varphi$. Another example can be easily obtained by fixing $n = 3$ so that:
\begin{equation}
	\textrm{d} \varphi \simeq \frac{\textrm{d} \phi}{\sqrt{1 + \lambda a^2 W_{\textrm{f}}^2 \phi^2 / 4 }} \ , \qquad \longrightarrow \qquad \varphi -\varphi_{\textrm{f}} = \frac{\sinh^{-1}(\sqrt{C} \phi)}{\sqrt{C}}  \ ,
\end{equation}
where again $C \equiv \lambda a^2 W_{\textrm{f}}^2 / 4 $. Once again it is possible to invert this equation and express $\phi$ as a function of $\varphi$.
\end{itemize}

Let us conclude this Appendix by expressing the lagrangian of DBI-like models (with $n=5/2$) in terms of the new field $\varphi$. By definition $(\textrm{d}\phi/\textrm{d}\varphi)^2 = 1 + C \phi $ and substituting Eq.~\eqref{varphi5/2} we can easily get:
\begin{equation}
	\left(\frac{\textrm{d}\phi}{\textrm{d}\varphi}\right)^2 = \left( \frac{C}{2} \right)^2 (\varphi  - \varphi_{\textrm{f}})^2 \ .
\end{equation}
We can then substitute into Eq.~\eqref{eq_non_standard:pDBI} to get:
\begin{equation}
	\label{eq_non_standard_varphi_p}
	p(X,\varphi) =  \frac{1}{\tilde{f}(\varphi)} \left[ 1 - \sqrt{1+2\tilde{f}(\varphi)  \left( \frac{C}{2} \right)^2 (\varphi -  \varphi_{\textrm{f}})^2 \, X } \ \right] - \tilde{V}(\varphi) \ ,
\end{equation}
where as $X \equiv g^{\mu\nu} \partial_{\mu} \varphi \partial_{\nu} \varphi /2$, $\tilde{f}(\varphi)\equiv f(\phi(\varphi))$ and $\tilde{V}(\varphi) \equiv V(\phi(\varphi)) $.

\section{Tachyonic inflation-like models.\label{sec:appendix_tachyon}}
Tachyon inflation is another model with a non-standard kinetic term arising in string theory~\cite{Gibbons:2002md,Padmanabhan:2002cp}. The slow-rolling scalar field leading to inflation is a tachyon\footnote{In the literature this field is typically denoted with $T$.}. The pressure in this model is given by:
\begin{align}
\label{eq_tachyon:pressure}
p(\phi,X)&=-V(\phi)\sqrt{1+2X}\;,
\end{align}
where as usual, $X\equiv-\frac{1}{2}\dot{\phi}^2$. It is useful to point out that in the case of tachyonic inflation the expression for the energy density is:
\begin{equation}
\label{eq_tachyon:energy}
	\rho \equiv 2Xp_{,X} - p = V(\phi) \left[  \frac{-2X}{\sqrt{1+2X}} +\sqrt{1+2X} \right] = \frac{V(\phi)}{\sqrt{1+2X}} = -p_{,X} \ ,
\end{equation}
so that the energy density is equal $-p_{,X}$.\\

The equation of state reads:
\begin{align}
\label{eq:tachyonicequationofstate}
\frac{\rho+p}{\rho}&=\dot{\phi}^2\;,
\end{align}
and the speed of sound:
\begin{align}
c_s^2&=1-\dot{\phi}^2\;.
\end{align}
The dynamics is obtained from the Friedmann equations \eqref{eq_non_standard:Friedmann}. An accelerated phase of expansion requires $\ddot{a}/a>0$ which implies that $\dot{\phi}^2$ is restricted to the range $\left[0,2/3\right]$. Note that the speed of sound moves away from its standard value during the phase of inflation and non-trivial non-Gaussianities are allowed. \\

This class of models can thus be described by the parametrization of Eq.~\eqref{eq_general:parametrization} fixing $\alpha=-1/2$, $f(\phi)=1$, $g(\phi)=-V(\phi)$. It is easy to show that in this case Eq.~\eqref{eq_general:phi_dot_param} thus reduces to:
	\begin{equation}
		-2X=\frac{W_{,\phi}^2}{V^2(\phi)}\left( 1 + 2X\right) \ .
	\end{equation}
Again we can express $X$ as a function of $\phi$ only:
	\begin{equation}
		\label{eq_tachyon:x_of_phi}
		-2X = \frac{W_{,\phi}^2/V^2(\phi)}{1 +  W_{,\phi}^2/V^2(\phi)} \ .
	\end{equation}
Substituting into Eq.~\eqref{eq_tachyon:energy} we thus get:
\begin{equation}
	\rho = -p_{,X} = \sqrt{V^2(\phi) +  W_{,\phi}^2} \ . 
\end{equation}
At this point we can use Eq.~\eqref{eq_formalism:friedman} to get:
\begin{equation}
	\label{eq_tachyon:potential}
	V^2(\phi) = \frac{9}{16} W^4 - W_{,\phi}^2  \ ,
\end{equation}
so that $-2X$ is directly given by:
	\begin{equation}
		\label{eq_tachyon:x_of_phi_last}
		-2X = \frac{W_{,\phi}^2}{\frac{9}{16} W^4 } \ .
	\end{equation}
Substituting into the expression for the $\beta$-function given in Eq.~\eqref{eq_formalism:beta_phi} we get:
\begin{equation}
	\label{eq_tachyon:beta_super}	
		\beta(\phi) = - \frac{4}{\sqrt{3}} \frac{W_{,\phi}}{W^2 } \ .
\end{equation}
The solution of this equation is thus given by:
	 \begin{equation}
	 	\label{eq_tachyon:superpot_beta}
	 	W(\phi) = W_{\textrm{f}} \left( 1 + W_{\textrm{f}} \int_{\phi_{\textrm{f}}}^{\phi} \frac{\sqrt{3}}{4} \beta(\hat{\phi}) \textrm{d}\hat{\phi} \right)^{-1} \ .
	 \end{equation}
We can then compare with Eq.~\eqref{eq_tachyon:potential} to get:
\begin{equation}
	V^2 = \frac{9}{16} W^4 \left( 1 - \frac{\beta^2}{3}\right) \ .
\end{equation}
We can then use Eq.~\eqref{eq_general:speed_of_sound} to compute the speed of sound:
\begin{equation}
	\label{eq_tachyonic:speed_of_sound}
	c_s^2 = \frac{1}{1+W_{,\phi}^2/V^2} \ .
\end{equation}
Similarly we can use Eq.~\eqref{eq_general:non_gaussianities} to show that:
\begin{equation}
	\label{eq_tachyonic:non_gaussianities}
	\frac{2\lambda}{\Sigma} = \frac{W_{,\phi}^2/V^2(\phi)}{1 +  W_{,\phi}^2/V^2(\phi)}  \ .
\end{equation}
As we are interested in describing inflation, which is realized for $\beta(\phi) \ll 1$, we can safely neglect $\beta^2/3$ with respect to one so that:
\begin{equation}
 	V^2 \simeq \frac{9}{16} W^4 \ .
 \end{equation} 
Under this assumption it is also possible to get:
\begin{equation}
	\label{eq_tachyon:beta_pot}	
		\beta(\phi) = - \frac{V_{,\phi}}{V^{3/2} } \ ,
\end{equation}
and also:
\begin{equation}
	c_s^2 \simeq \frac{1}{1+\beta^2 / 3} \ , \qquad \frac{2\lambda}{\Sigma} = \frac{\beta^2 }{3 +  \beta^2}  \ .
\end{equation}
As a consequence for $\beta^2 \rightarrow 0$ we both have $c_s^2 \rightarrow 1$ and negligible non-Gaussianities. Actually inflation can also be realized for $\beta \simeq - C$ with $C $ sufficiently small\footnote{As discussed in~\cite{Binetruy:2014zya} this case actually corresponds to the case of power-law inflation~\cite{Lucchin:1984yf}.}. In this case we have:
	\begin{equation}
	 	W(\phi) = \frac{W_{\textrm{f}}}{1 + W_{\textrm{f}} C \phi} \ ,
	\end{equation} 	
$c_s^2$ approaches a constant value different both from zero and one and $2\lambda / \Sigma$ approaches a constant non-zero value.\\

We conclude this Appendix by considering some particular parameterizations for the $\beta$-function. In particular we show that by considering some simple expressions for the $\beta$-function we can recover the models presented in~\cite{Steer:2003yu}.\\

For example we can consider the case $\beta(\phi) = a \phi^n$. Using Eq.~\eqref{eq_tachyon:superpot_beta} we get:
	\begin{equation}
	 	\label{eq_tachyon:superpot_monom}
	 	W(\phi) = W_{\textrm{f}} \left( 1 + a W_{\textrm{f}} \frac{\sqrt{3}}{4} \frac{\phi^{m+1}}{m+1} \right)^{-1} \ ,
	 \end{equation}
that corresponds to the case of inverse power-law potential of~\cite{Steer:2003yu,Brax:2003rs,Abramo:2003cp}.\\

Similarly we can fix other parameterizations of the $\beta$-function in order to recover the Exponential potential of~\cite{Steer:2003yu,Sami:2002fs,Sen:2002an} or the Inverse cosh potential of~\cite{Steer:2003yu,Kim:2003he}.

\providecommand{\href}[2]{#2}\begingroup\raggedright


\begin{thebibliography}{10}



\bibitem{Binetruy:2014zya} 
  P.~Binetruy, E.~Kiritsis, J.~Mabillard, M.~Pieroni and C.~Rosset,
  JCAP {\bf 1504}, no. 04, 033 (2015)
  doi:10.1088/1475-7516/2015/04/033
  [arXiv:1407.0820 [astro-ph.CO]].
  

\bibitem{McFadden:2009fg} 
  P.~McFadden and K.~Skenderis,
  Phys.\ Rev.\ D {\bf 81}, 021301 (2010)
  doi:10.1103/PhysRevD.81.021301
  [arXiv:0907.5542 [hep-th]].

\bibitem{McFadden:2010na} 
  P.~McFadden and K.~Skenderis,
  J.\ Phys.\ Conf.\ Ser.\  {\bf 222}, 012007 (2010)
  doi:10.1088/1742-6596/222/1/012007
  [arXiv:1001.2007 [hep-th]].


\bibitem{Garriga:2014fda} 
  J.~Garriga, K.~Skenderis and Y.~Urakawa,
  JCAP {\bf 1501}, no. 01, 028 (2015)
  doi:10.1088/1475-7516/2015/01/028
  [arXiv:1410.3290 [hep-th]].

\bibitem{Garriga:2014ema} 
  J.~Garriga and Y.~Urakawa,
  JHEP {\bf 1406}, 086 (2014)
  doi:10.1007/JHEP06(2014)086
  [arXiv:1403.5497 [hep-th]].

\bibitem{Garriga:2015tea} 
  J.~Garriga, Y.~Urakawa and F.~Vernizzi,
  JCAP {\bf 1602}, no. 02, 036 (2016)
  doi:10.1088/1475-7516/2016/02/036
  [arXiv:1509.07339 [hep-th]].

\bibitem{Kiritsis:2013gia} 
  E.~Kiritsis,
  JCAP {\bf 1311}, 011 (2013)
  doi:10.1088/1475-7516/2013/11/011
  [arXiv:1307.5873 [hep-th]].


\bibitem{Pieroni:2016gdg} 
  M.~Pieroni,
  arXiv:1611.03732 [gr-qc].

\bibitem{ArmendarizPicon:1999rj} 
  C.~Armendariz-Picon, T.~Damour and V.~F.~Mukhanov,
  Phys.\ Lett.\ B {\bf 458}, 209 (1999)
  doi:10.1016/S0370-2693(99)00603-6
  [hep-th/9904075].

\bibitem{Silverstein:2003hf} 
  E.~Silverstein and D.~Tong,
  Phys.\ Rev.\ D {\bf 70}, 103505 (2004)
  doi:10.1103/PhysRevD.70.103505
  [hep-th/0310221].

\bibitem{Alishahiha:2004eh} 
  M.~Alishahiha, E.~Silverstein and D.~Tong,
  Phys.\ Rev.\ D {\bf 70}, 123505 (2004)
  doi:10.1103/PhysRevD.70.123505
  [hep-th/0404084].


\bibitem{Gibbons:2002md} 
  G.~W.~Gibbons,
  Phys.\ Lett.\ B {\bf 537}, 1 (2002)
  doi:10.1016/S0370-2693(02)01881-6
  [hep-th/0204008].

\bibitem{Padmanabhan:2002cp} 
  T.~Padmanabhan,
  Phys.\ Rev.\ D {\bf 66}, 021301 (2002)
  doi:10.1103/PhysRevD.66.021301
  [hep-th/0204150].

\bibitem{Steer:2003yu} 
  D.~A.~Steer and F.~Vernizzi,
  Phys.\ Rev.\ D {\bf 70}, 043527 (2004)
  doi:10.1103/PhysRevD.70.043527
  [hep-th/0310139].

\bibitem{Li:2013cem} 
  S.~Li and A.~R.~Liddle,
  JCAP {\bf 1403}, 044 (2014)
  doi:10.1088/1475-7516/2014/03/044
  [arXiv:1311.4664 [astro-ph.CO]].


\bibitem{Salopek:1990jq} 
  D.~S.~Salopek and J.~R.~Bond,
  Phys.\ Rev.\ D {\bf 42}, 3936 (1990).
  doi:10.1103/PhysRevD.42.3936


\bibitem{Pieroni:2015cma} 
  M.~Pieroni,
  JCAP {\bf 1602}, no. 02, 012 (2016)
  doi:10.1088/1475-7516/2016/02/012
  [arXiv:1510.03691 [hep-ph]].


\bibitem{Maldacena:2002vr} 
  J.~M.~Maldacena,
  JHEP {\bf 0305}, 013 (2003)
  doi:10.1088/1126-6708/2003/05/013
  [astro-ph/0210603].

\bibitem{Weinberg:2005vy} 
  S.~Weinberg,
  Phys.\ Rev.\ D {\bf 72}, 043514 (2005)
  doi:10.1103/PhysRevD.72.043514
  [hep-th/0506236].

\bibitem{Chen:2006nt} 
  X.~Chen, M.~x.~Huang, S.~Kachru and G.~Shiu,
  JCAP {\bf 0701}, 002 (2007)
  doi:10.1088/1475-7516/2007/01/002
  [hep-th/0605045].

\bibitem{Chen:2010xka} 
  X.~Chen,
  Adv.\ Astron.\  {\bf 2010}, 638979 (2010)
  doi:10.1155/2010/638979
  [arXiv:1002.1416 [astro-ph.CO]].

\bibitem{Acquaviva:2002ud} 
  V.~Acquaviva, N.~Bartolo, S.~Matarrese and A.~Riotto,
  Nucl.\ Phys.\ B {\bf 667}, 119 (2003)
  doi:10.1016/S0550-3213(03)00550-9
  [astro-ph/0209156].

\bibitem{Babich:2004gb} 
  D.~Babich, P.~Creminelli and M.~Zaldarriaga,
  JCAP {\bf 0408}, 009 (2004)
  doi:10.1088/1475-7516/2004/08/009
  [astro-ph/0405356].

\bibitem{Ade:2015xua} 
  P.~A.~R.~Ade {\it et al.} [Planck Collaboration],
  Astron.\ Astrophys.\  {\bf 594}, A13 (2016)
  doi:10.1051/0004-6361/201525830
  [arXiv:1502.01589 [astro-ph.CO]].

\bibitem{Ade:2015lrj} 
  P.~A.~R.~Ade {\it et al.} [Planck Collaboration],
  doi:10.1051/0004-6361/201525898
  arXiv:1502.02114 [astro-ph.CO].

\bibitem{Ade:2015ava} 
  P.~A.~R.~Ade {\it et al.} [Planck Collaboration],
  doi:10.1051/0004-6361/201525836
  arXiv:1502.01592 [astro-ph.CO].


\bibitem{Schwarz:2001vv} 
  D.~J.~Schwarz, C.~A.~Terrero-Escalante and A.~A.~Garcia,
  Phys.\ Lett.\ B {\bf 517}, 243 (2001)
  doi:10.1016/S0370-2693(01)01036-X
  [astro-ph/0106020].

\bibitem{Lucchin:1984yf} 
  F.~Lucchin and S.~Matarrese,
  Phys.\ Rev.\ D {\bf 32}, 1316 (1985).
  doi:10.1103/PhysRevD.32.1316

\bibitem{Brax:2003rs} 
  P.~Brax, J.~Mourad and D.~A.~Steer,
  Phys.\ Lett.\ B {\bf 575}, 115 (2003)
  doi:10.1016/j.physletb.2003.09.039
  [hep-th/0304197].

\bibitem{Abramo:2003cp} 
  L.~R.~W.~Abramo and F.~Finelli,
  Phys.\ Lett.\ B {\bf 575}, 165 (2003)
  doi:10.1016/j.physletb.2003.09.065
  [astro-ph/0307208].

\bibitem{Sami:2002fs} 
  M.~Sami, P.~Chingangbam and T.~Qureshi,
  Phys.\ Rev.\ D {\bf 66}, 043530 (2002)
  doi:10.1103/PhysRevD.66.043530
  [hep-th/0205179].

\bibitem{Sen:2002an} 
  A.~Sen,
  Mod.\ Phys.\ Lett.\ A {\bf 17}, 1797 (2002)
  doi:10.1142/S0217732302008071
  [hep-th/0204143].
  
\bibitem{Kim:2003he} 
  C.~j.~Kim, H.~B.~Kim, Y.~b.~Kim and O.~K.~Kwon,
  JHEP {\bf 0303}, 008 (2003)
  doi:10.1088/1126-6708/2003/03/008
  [hep-th/0301076].

\bibitem{Bourdier:2013axa} 
  J.~Bourdier and E.~Kiritsis,
  Class.\ Quant.\ Grav.\  {\bf 31}, 035011 (2014)
  doi:10.1088/0264-9381/31/3/035011
  [arXiv:1310.0858 [hep-th]].

\bibitem{Germani:2010gm} 
  C.~Germani and A.~Kehagias,
  Phys.\ Rev.\ Lett.\  {\bf 105}, 011302 (2010)
  doi:10.1103/PhysRevLett.105.011302
  [arXiv:1003.2635 [hep-ph]].

\bibitem{Germani:2010ux} 
  C.~Germani and A.~Kehagias,
  JCAP {\bf 1005}, 019 (2010)
  Erratum: [JCAP {\bf 1006}, E01 (2010)]
  doi:10.1088/1475-7516/2010/05/019, 10.1088/1475-7516/2010/06/E01
  [arXiv:1003.4285 [astro-ph.CO]].

  
\end{thebibliography}
\end{document}